\documentclass[preprint,review,12pt]{elsarticle}

\usepackage{amssymb}
\usepackage{bm}

\journal{arXiv}

\begin{document}

\begin{frontmatter}



\title{Aging prediction using deep generative model toward the development of preventive medicine}

\author[inst1]{Hisaichi Shibata\corref{cor1}}
\cortext[cor1]{Corresponding Author}
\ead{sh@g.ecc.u-tokyo.ac.jp}
\affiliation[inst1]{organization={Department of Radiology, The~University of Tokyo Hospital},
            addressline={7-3-1~Hongo}, 
            city={Bunkyo, Tokyo},
            postcode={113-8655},
            country={Japan}}
\affiliation[inst2]{organization={Department of Computational Diagnostic Radiology and Preventive Medicine, The~University of Tokyo Hospital},
            addressline={7-3-1~Hongo}, 
            city={Bunkyo, Tokyo},
            postcode={113-8655}, 
            country={Japan}}
\author[inst1]{Shouhei Hanaoka}
\author[inst2,inst3]{Yukihiro Nomura}
\author[inst2]{Naoto~Hayashi}
\author[inst1]{Osamu~Abe}
\affiliation[inst3]{organization={Center for Frontier Medical Engineering, Chiba~University},
            addressline={1-33~Yayoi-cho}, 
            city={Inage, Chiba},
            postcode={263-8522}, 
            country={Japan}}

\begin{abstract}
From birth to death, we all experience surprisingly ubiquitous changes over time due to aging.
If we can predict aging in the digital domain, that is, the digital twin of the human body, we would be able to detect lesions in their very early stages, thereby enhancing the quality of life and extending the life span.
We observed that none of the previously developed digital twins of the adult human body explicitly trained longitudinal conversion rules between volumetric medical images with deep generative models, potentially resulting in poor prediction performance of, for example, ventricular volumes.
Here, we establish a new digital twin of an adult human body that adopts longitudinally acquired head computed tomography (CT) images for training, enabling prediction of future volumetric head CT images from a single present volumetric head CT image.
We, for the first time, adopt one of the three-dimensional flow-based deep generative models to realize this sequential three-dimensional digital twin.
We show that our digital twin outperforms the latest methods of prediction of ventricular volumes in relatively short terms.
\end{abstract}




\end{frontmatter}


\section{Introduction}
From birth to death, we all experience surprisingly ubiquitous changes over time due to aging.
Here, we establish a model that can predict aging in the digital domain, that is, a digital twin of the human body.
With this digital twin, we can improve preventive medicine, thereby enhancing the quality of life and prolonging life.
Specifically, we can detect lesions in their very early stages.
Ultimately, we can predict lesions before their occurrence. 

We adopt a flow-based deep generative model (DGM), which is one of DGMs \cite{kingma2018glow,shibata2021versatile,dinh2014nice, dinh2016density, goodfellow2014generative, kingma2013auto} to realize the digital twin.
DGMs can learn a probabilistic distribution of data, e.g., images, with deeply stacked neural networks.
DGMs automatically find rules in images; hence, we can model and predict aging, which involves discrete and probabilistic changes. 

Several authors developed digital twins of the human body that can predict aging and disease progression\cite{choi2018predicting, rachmadi2020automatic, peng2021longitudinal,xia2021learning,ravi2022degenerative}.
However, previous studies are limited to modeling of brain magnetic resonance (MR) or positron emission tomography (PET) images.
Critically, we observed that in previous studies, digital twins of an adult human body did not explicitly learn longitudinal conversion rules between volumetric medical images with DGMs, potentially resulting in poor prediction performance of, for example, ventricular volumes.
Moreover, none of the previous studies of digital twins of an adult human body adopted a three-dimensional (3D) convolutional network. (We discuss the structure of neural networks in the next section.) 

In this study, we aim to establish a digital twin of an adult human body.
By learning longitudinal head computed tomography (CT) images, the digital twin can predict future ventricular volumes from a single present volumetric head CT image.
Moreover, we show that this digital twin can more precisely predict ventricular volume than previous digital twins in relatively short terms.
We develop a conditional model of CT images for healthy adult people to simplify this problem.
Still, we can potentially extend our conditional model to the unconditional modeling of aging for any portion of the human body.

\section{Related works}
In this section, we briefly review previous studies on aging prediction using DGMs.
Choi et al. \cite{choi2018predicting} developed the first digital twin for aging prediction adopting a variational autoencoder (VAE \cite{kingma2013auto})-based network trained with cross-sectional 3D brain PET images.
Rachmadi et al. \cite{rachmadi2020automatic} developed the first digital twin for aging prediction adopting generative adversarial network (GAN \cite{goodfellow2014generative})-based deep neural networks trained with longitudinal two-dimensional (2D) brain MR images.
Peng et al. \cite{peng2021longitudinal} developed the first digital twin for aging prediction adopting GAN-based deep neural networks trained with longitudinal 3D brain MR images of infants.
Xia et al. \cite{xia2021learning} developed the first digital twin for aging prediction adopting GAN-based deep neural networks trained with cross-sectional 2D brain MR images.
Their digital twin is for 2D images, but they can handle 3D images by decomposing 3D images into 2D slice images and inputting those 2D slices into the digital twin.
Because their digital twin learns from cross-sectional data and contains only a 2D network, the accuracy of aging prediction may be affected.
Finally, the spatial resolution of their digital twin is 208$\times$150.
Ravi et al. \cite{ravi2022degenerative} developed the first digital twin for aging prediction adopting GAN-based deep neural networks trained with cross-sectional 3D brain MR images.
Although the final spatial resolution of a generated volumetric image is not explicitly stated in their paper, their digital twin contains a 3D super-resolution network (up-sampling four times) in the final stage.
The digital twin can handle 3D images, but the digital twin combines only 2D layers to realize it; hence, this can decrease the accuracy of aging prediction.

In this study, we adopted a generator that contains full 3D convolutional neural networks; hence, our digital twin can naturally handle 3D images.
We, for the first time, realize aging prediction using longitudinally acquired 3D medical images with a DGM.
In contrast to all previous studies, we used head CT images instead of MR or PET images. 
The spatial resolution of our digital twin is $128^3$, and we realized it without using super-resolution blocks.
Moreover, we can straightly extend the application of our digital twin to images acquired with other modalities for any part of the human body.
In a previous study (\cite{ravi2022degenerative}), 9,852 volumetric MR images were used for its training, but we used only 1,024 unique volumetric CT images.

\section{Materials}
The ethical review board of our institution approved this study.
All the subjects gave informed consent to use their CT images.
We used 1,051 head CT images.
A GE LightSpeed CT scanner (GE Healthcare, Waukesha, WI, USA) at our institution scanned these images.
The acquisition parameters were as follows: number of detector rows, 16; tube voltage, 140 kVp; tube current, 120 mA; rotation time, 0.5 s; moving table speed, 21.9 mm/s; body filter, soft; reconstruction slice thickness and interval, 1.25 mm; field of view, 500 mm; matrix size, 512$\times$512 pixels; pixel spacing, 0.977 mm.
We randomly divided the head CT images into training (1,024) and test datasets (27).
Because we manually evaluated one of the prediction performances for aging, we set the number of head CT images in the test dataset to be small.
We converted the 16-bit head CT images $I_\mathrm{src}$ (CT number in HU) into 8-bit images $I_\mathrm{dst}$ (similar to an image in the brain window) with the following empirical formula:
\begin{eqnarray}
    I_\mathrm{dst} =  \mathrm{clip}\left(I_\mathrm{src} + 80, 0, 255\right),
\end{eqnarray}
where the operator $\mathrm{clip}(x, a, b)$ restricts the value range of an array $x$ from $a$ to $b$.
After this conversion for each image, we computed the center of gravity, then extracted the head region ($256^3$).
Owing to limits in GPU memory (48 GB), we downsampled these images into a resolution of $128^3$; hence, we set $D=H=W=128$ and $C=1$.

\section{Methods}
We predict the volumetric CT image of a future human body ($\bm{x}^+ \in \mathbf{R}^{D\times H\times W \times C}$) from a single volumetric CT image of a previous human body ($\bm{x}^- \in \mathbf{R}^{D\times H\times W \times C}$).
We divided this problem into two: the spatial model and temporal model of volumetric CT images.

\subsection{Spatial model}
We adopted 3D-GLOW \cite{shibata2021versatile}, which is one of the flow-based DGMs \cite{dinh2014nice,dinh2016density,kingma2018glow}.
We can adopt other DGMs, e.g., VAEs, GANs, and diffusion models, and their hybrids.
However, typical VAEs generate blurred images.
Typical diffusion models require relatively higher computational costs in training and generation of images.
Typical GANs suffer from mode collapse.
Moreover, we must embed real images on the latent space of a GAN.   
Some research groups proposed this embedding method for natural images.
However, we must embed medical images, which require additional human resources.
Although our 3D-GLOW is difficult to train with high-resolution images, we established a novel progressive learning \cite{shibata2021x2ct}, which is a kind of curriculum learning to improve it.

The flow-based DGMs can explicitly estimate the probability density function for images, i.e., density estimation, unlike GANs and VAEs.
Moreover, the flow-based DGMs simultaneously learn encoder and decoder networks, i.e., flow; hence, we do not have to prepare an encoder network to embed a real image onto the latent space unlike GANs.

For given images ($\bm{x}_i \in \mathcal{D}$, where the subscript $i$ distinguishes images), the flow-based DGMs directly minimize the averaged negative logarithm likelihoods (NLL, a lower value is better),  $\frac{1}{|\mathcal{D}|}\sum^{\mathcal{D}} -\log p (\bm{x}_i)$.
By adopting the change-of-variable formula, we convert the above logarithm likelihood of intractable probabilistic distribution for given images to the logarithm likelihood of tractable probabilistic distribution, i.e., an element-wise independent normal distribution, as follows:
\begin{eqnarray}
    \log p (\bm{x}_i) = \log p (\bm{z}_i) + \log \left| \det \left(\frac{\partial \bm{z}_i}{\partial \bm{x}_i} \right)\right|,
\end{eqnarray}
where $\bm{z}_i \sim \mathcal{N}\left( \bm{\mu}, \bm{\sigma}^2 \right)$ and we call $\bm{z}_i$ the latent space vector.
We train the means $\bm{\mu}$ and the variances $\bm{\sigma}^2$ with the flow-based DGMs from given images.
To handle the second term, we put $\bm{z}_i = \bm{G}_{\bm{\theta}} \left(\bm{x}_i\right)$, where $\bm{G}_{\bm{\theta}}$ is an invertible vector function with trainable parameters $\bm{\theta}$.
To efficiently compute the determinant, we usually decompose $\bm{G}_{\bm{\theta}}$ into a composition of invertible vector functions.

\begin{figure}
    \centering
    \includegraphics[width=12cm]{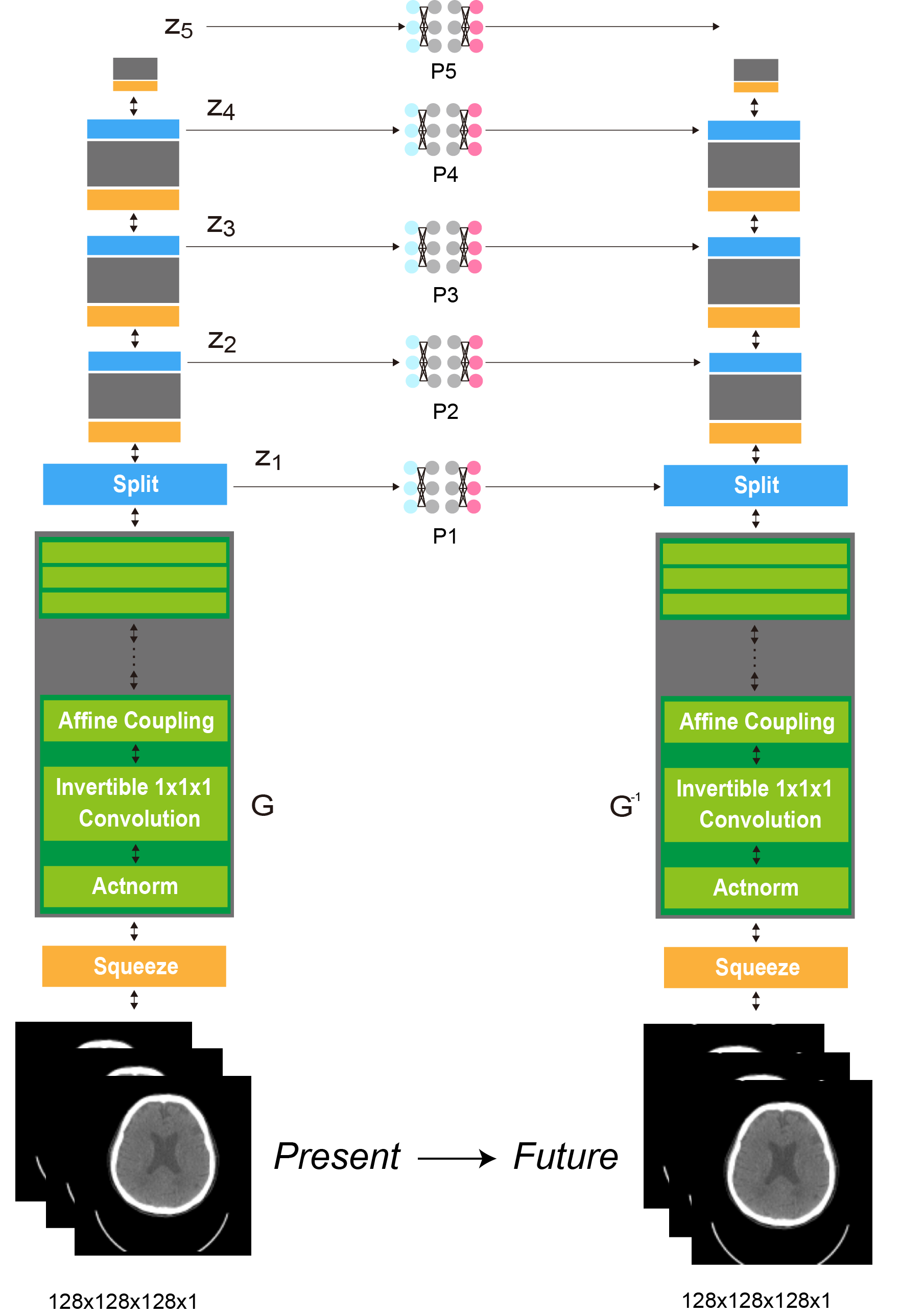}
    \caption{Architecture of our digital twin.}
    \label{fig:digital_twin_arch}
\end{figure}

To efficiently learn the probability density function for images, GLOW \cite{kingma2018glow} and 3D GLOW \cite{shibata2021versatile} adopt a multiscale architecture.
The architecture outputs multiple latent space vectors (e.g., $\bm{z}_1$ to $\bm{z}_5$) in different levels of a neural network (see Fig.~\ref{fig:digital_twin_arch} and \cite{kingma2018glow} for details).
Owing to the multiscale architecture of 3D GLOW, we distinguish the latent space vector $\bm{z}_i$ in different layer levels using the subscript $l (=1,2,3,4,5)$, e.g., $\bm{z}_{i,l}$.

\begin{table}[htb]
  \begin{center}
    \caption{Hyperparameters used to train 3D GLOW model.}
    \label{tab:hps}    
    \begin{tabular}{lc} \hline
      Flow coupling & Affine \\
      Learn-top option & True \\
      Flow permutation & 1$\times$1$\times$1 convolution \\
      Minibatch size & 1 per GPU \\
      Train epochs & 70 (1 bit *1) \\
      & 30 (2 bit from 1 bit *1) \\
      & 10 (3 bit from 2 bit *1) \\      
      & 20 (4 bit from 3 bit *1) \\ 
      & 10 (5 bit from 4 bit *1) \\
      & 10 (6 bit from 5 bit *1) \\    
      & 10 (7 bit from 6 bit *1) \\ 
      & 200 (8 bit from 7 bit *2) \\                
      Layer levels & 5 \\
      Depth per level & 8 \\
      Filter width & 512 \\
      Learning rate in steady state & $1.0 \times 10^{-3}$ (*1) \\
                                & $1.0 \times 10^{-4}$ (*2) \\      
      Epochs warm up & 100 \\
      \hline
    \end{tabular}
    \end{center}
\end{table}
Table~\ref{tab:hps} shows the hyperparameters used to train 3D-GLOW.
We trained 3D-GLOW in a progressive manner \cite{shibata2021x2ct}.

\subsection{Temporal model}
We adopt other deep neural networks to convert $\bm{z}^-$ into $\bm{z}^+$.
We can mathematically express this conversion as
\begin{eqnarray}
    \bm{z}^+ = \bm{P}_{\bm{\phi}} \left( \bm{z}^- \right),
\end{eqnarray}
and we train parameters $\bm{\phi}$. The total dimension of $\bm{z}^\pm$ is $128^3$.

We model the conversion for each layer level.
Therefore, for $1 \le l\le 5$, we have
\begin{eqnarray}
    \bm{z}^+_l = \bm{P}_{l, \bm{\phi}_l} \left( \bm{z}^-_l \right),
\end{eqnarray}
where $\bm{\phi}_l$ represents parameters in a deep neural network $\bm{P}_{l}$ for a latent space of each layer level. 
We designed $\bm{P}_{l}$ as follows:
\begin{eqnarray}
    \bm{P}_1 = \mathrm{Conv3D(filters=4, kernel\_size=3,}\nonumber  \\
    \mathrm{padding=same, activation=relu)} + I, \\
    \bm{P}_2 = [\mathrm{Conv3D(filters=16, kernel\_size=3,} \nonumber\\ \mathrm{padding=same, activation=relu)} \nonumber \\
      \circ \mathrm{Conv3D(filters=16, kernel\_size=3,}\nonumber\\ \mathrm{padding=same, activation=relu)}] + I, \\
      \bm{P}_3 = [\mathrm{Conv3D(filters=64, kernel\_size=3,} \nonumber \\ \mathrm{padding=same, activation=relu)} \nonumber \\
      \circ \mathrm{Conv3D(filters=64,  kernel\_size=3, }\nonumber \\
      \mathrm{padding=same, activation=relu)} \nonumber\\
      \circ \mathrm{Conv3D(filters=64,  kernel\_size=3, }\nonumber \\
      \mathrm{padding=same, activation=relu)} \nonumber\\      
      \circ \mathrm{Conv3D(filters=64,  kernel\_size=3, }\nonumber \\
      \mathrm{padding=same, activation=relu)}] + I, \\
      \bm{P}_4 = [\mathrm{Conv3D(filters=256,  kernel\_size=3,}\nonumber \\
      \mathrm{padding=same, activation=relu)} \nonumber \\
      \circ \mathrm{Conv3D(filters=256,  kernel\_size=3, }\nonumber \\
      \mathrm{padding=same, activation=relu)} \nonumber\\
      \circ \mathrm{Conv3D(filters=256,  kernel\_size=3,}\nonumber \\
      \mathrm{padding=same, activation=relu)} \nonumber\\      
      \circ \mathrm{Conv3D(filters=256, kernel\_size= 3,}\nonumber \\
      \mathrm{padding=same, activation=relu)}] + I, \\  
      \bm{P}_5 = [\mathrm{Conv3D(filters=2048, kernel\_size= 3,}\nonumber \\
      \mathrm{padding=same, activation=relu)} \nonumber \\
      \circ \mathrm{Conv3D(filters=2048, kernel\_size= 3,}\nonumber \\
      \mathrm{padding=same, activation=relu)} \nonumber\\
      \circ \mathrm{Conv3D(filters=2048,  kernel\_size=3,}\nonumber \\
      \mathrm{padding=same, activation=relu)} \nonumber\\      
      \circ \mathrm{Conv3D(filters=2048,  kernel\_size=3,}\nonumber \\
      \mathrm{padding=same, activation=relu)}] + I,      
\end{eqnarray}
where $I$ represents the identity function.
The above notation is from Keras documentation \cite{chollet2015keras}.

We did not input the raw latent space vector.
Before the input, we normalized the latent space vector $\bm{z}^-$ and $\bm{z}^+$ as follows:
\begin{eqnarray}
    \bm{z}^- \leftarrow \mathrm{clip}\left(\frac{\bm{z}^- + a}{b}, 0, 1\right), \\
    \bm{z}^+ \leftarrow \mathrm{clip}\left(\frac{\bm{z}^+ + a}{b}, 0, 1\right),
\end{eqnarray}
where $a(=24)$ and $b(=48)$ are parameters; we empirically determined $a$ and $b$ assuming that almost all of the elements in the latent space vector distribute in $-a<z<a$.

To minimize the mean squared error for the predicted volumetric image and the ground-truth image expressed as the latent space vector in the training dataset, we trained the neural networks ($\bm{P}_1$ to $\bm{P}_5$) using a stochastic gradient decent (SGD) method with the learning rate $(=10.0)$ for 20 epochs.
The number of pairs of $\bm{z}^\pm$ for the training of the temporal model was 640 (including duplication of subjects), which are randomly depicted from the training dataset for the spatial model. 

\subsection{Spatiotemporal model}
First, we train 3D-GLOW using many volumetric CT images.
Then, we obtain an invertible generator ($\bm{G}_{\bm{\theta}}$), that can generate fake but realistic volumetric CT images from a noise vector.  
Second, we map previous volumetric CT images ($\bm{x}^-$) and next volumetric CT images ($\bm{x}^+$) into the latent space of 3D-GLOW ($\bm{z}^-$ and $\bm{z}^+$) by $\bm{G}_{\bm{\theta}}$.
Then, we train temporal conversion rules between $\bm{z}^-$ and $\bm{z}^+$ with the other deep neural networks ($\bm{P}_{\bm{\phi}}$).
Finally, we map a present volumetric CT image onto the latent space, predicting the latent space expression of a future volumetric CT image.
Then, we map the future latent space expression into a future volumetric CT image by $\bm{G}^{-1}_{\bm{\theta}}$.

To summarize, we have
\begin{eqnarray}
    \bm{x}^+ &=& \mathcal{M} \left( \bm{x}^- \right), \\
    \mathcal{M} &:=& \bm{G}^{-1}_{\bm{\theta}} \circ \bm{P}_{\bm{\phi}} \circ \bm{G}_{\bm{\theta}},
\end{eqnarray}
where $\mathcal{M}$ is an operator to step the time of an image.
We can predict the future aging ($N$ times further) by recursively acting the operator $\mathcal{M}$:
\begin{eqnarray}
    \underbrace{\mathcal{M} \circ \cdots \circ \mathcal{M}}_{N \mathrm{-times}} &=&  \bm{G}^{-1}_{\bm{\theta}} \circ \bm{P}_{\bm{\phi}} \circ \bm{G}_{\bm{\theta}} \circ \cdots \circ  \bm{G}^{-1}_{\bm{\theta}} \circ \bm{P}_{\bm{\phi}} \circ \bm{G}_{\bm{\theta}} \nonumber \\ 
    &=& \bm{G}^{-1}_{\bm{\theta}} \circ \left( \bm{P}_{\bm{\phi}}\right)^{N} \circ \bm{G}_{\bm{\theta}}.
\end{eqnarray}

We separately trained parameters $\bm{\theta}$ and $\bm{\phi}$.
We utilized Tensorflow 1.14.0 for the back end of the deep neural networks.
The CUDA and cuDNN versions used were 10.0.130 and 7.4, respectively.
We carried out all processes on a single workstation consisting of two Intel Xeon Gold 6230 processors, 384 GB memory, and five GPUs (NVIDIA Quadro RTX 8000 with 48 GB of memory).
For the training of 3D-GLOW, we used only four GPUs out of the five GPUs, and for other processes, we utilized only one GPU.

\subsection{Quantification}
A radiologist semi-automatically segmented the ventricular volumes of brains using Simple ITK \cite{beare2018image}.
We computed ventricular volumes from the segmentation results.

\section{Results}

\begin{figure}
    \centering
    \includegraphics[width=12cm]{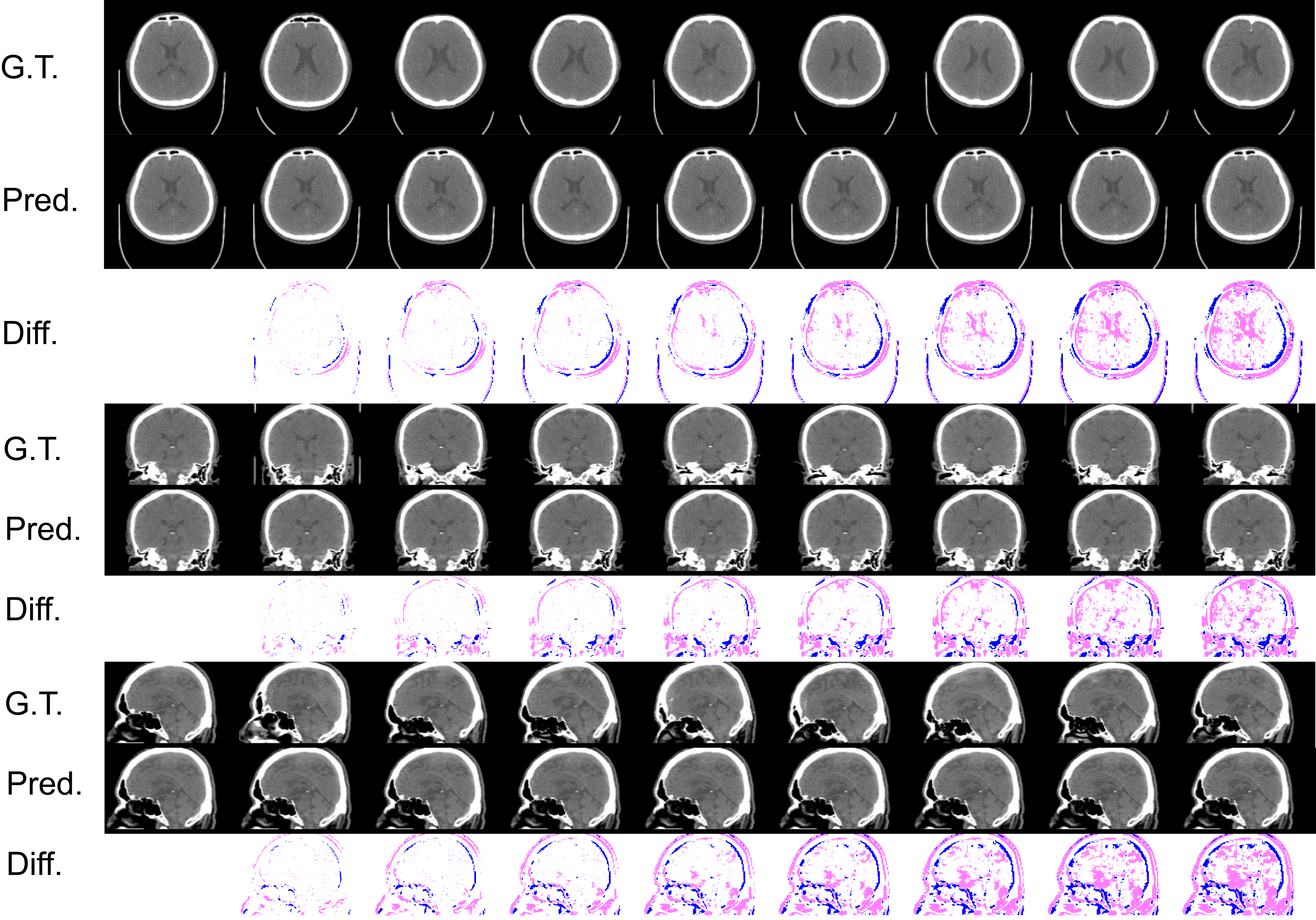}
    \caption{Result of aging prediction (case 1); from left to right, 0 to 8 years.}
    \label{fig:pred_case1}
\end{figure}
\begin{figure}
    \centering
    \includegraphics[width=12cm]{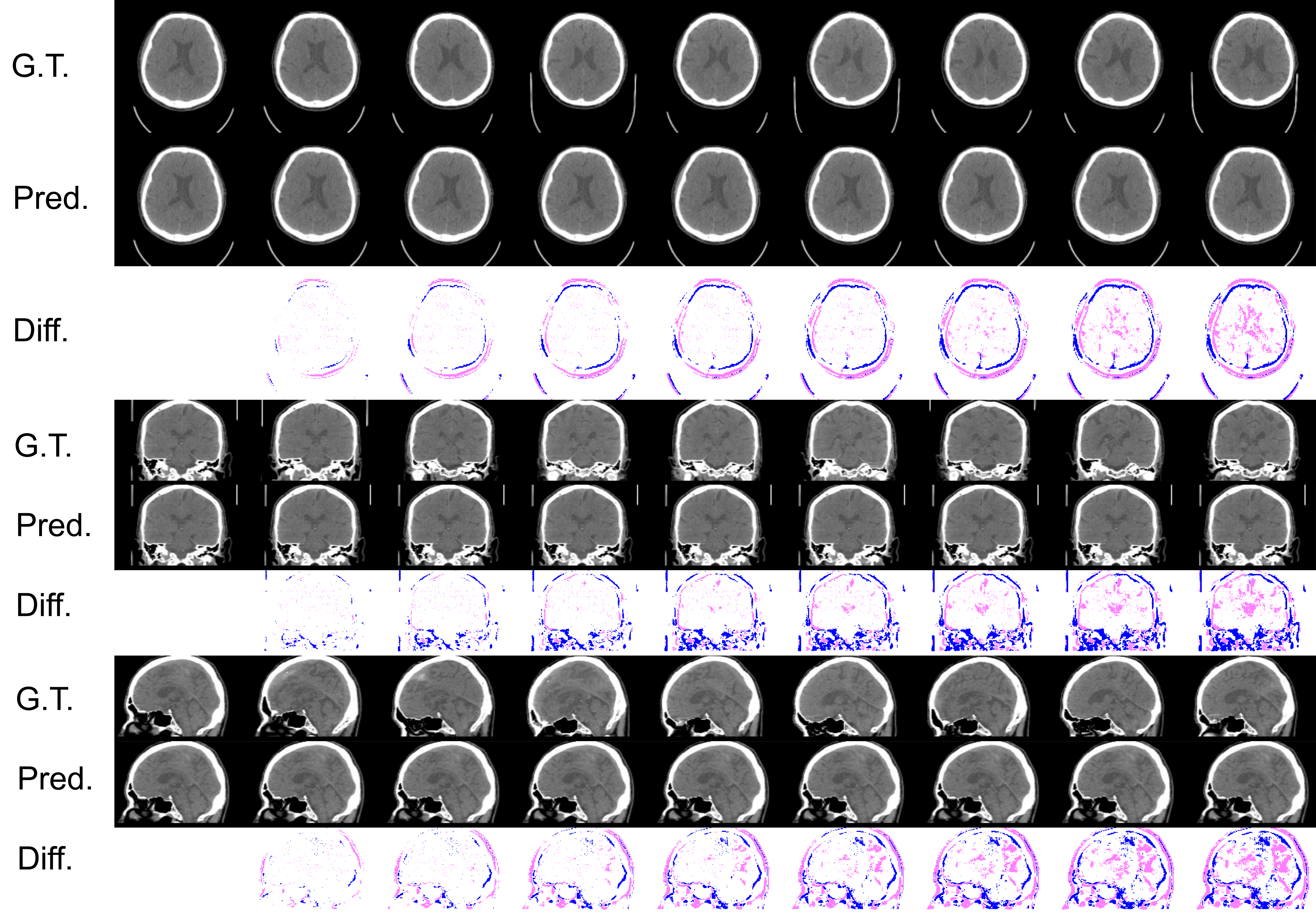}
    \caption{Result of aging prediction (case 2); from left to right, 0 to 8 years.}
    \label{fig:pred_case2}
\end{figure}
\begin{figure}
    \centering
    \includegraphics[width=12cm]{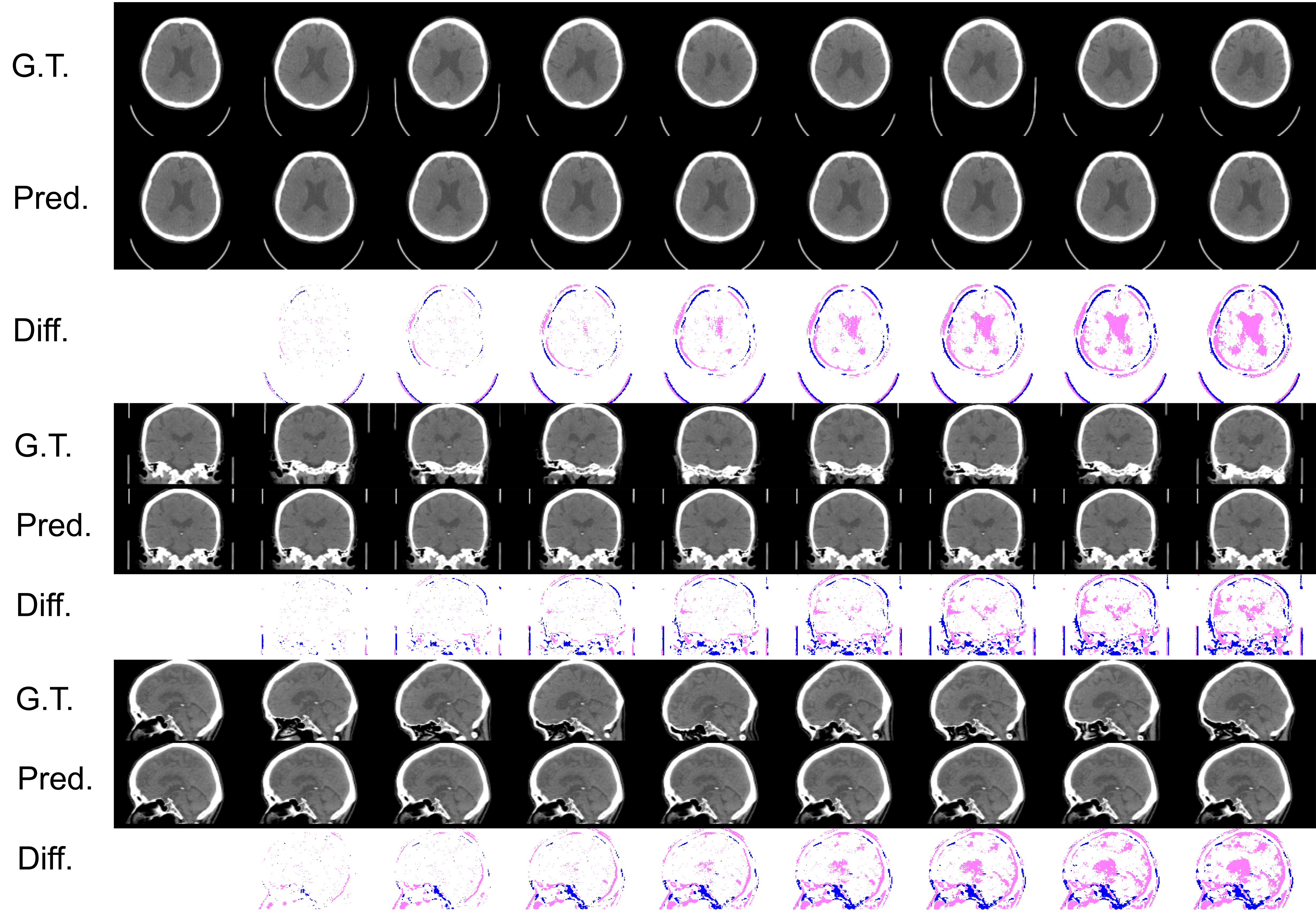}
    \caption{Result of aging prediction (case 3); from left to right, 0 to 8 years.}
    \label{fig:pred_case3}
\end{figure}

\begin{figure}
    \centering
    \includegraphics[width=8cm]{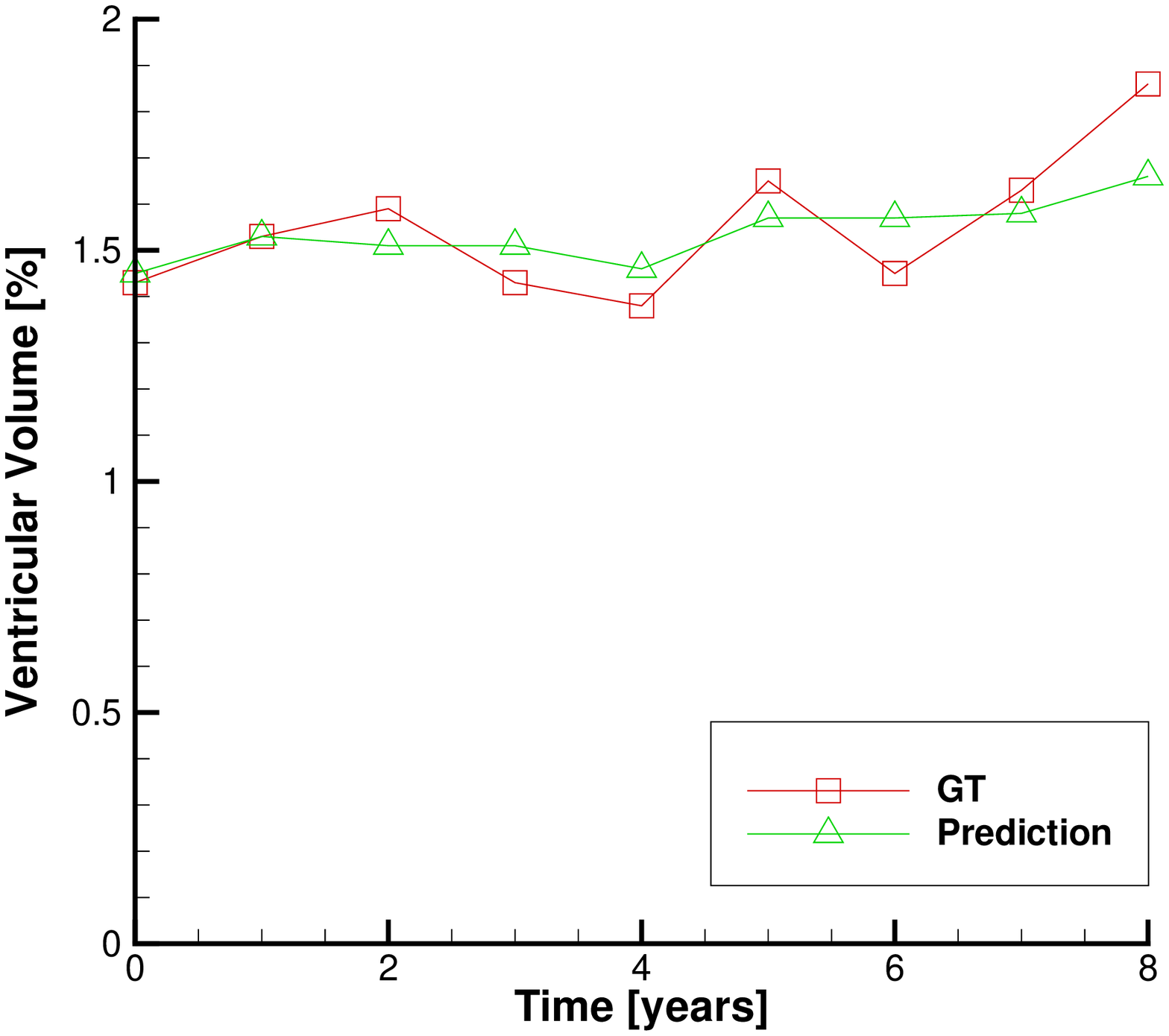}
    \caption{Ground truth and prediction results for ventricular volumes (case 1).}
    \label{fig:case1}
\end{figure}
\begin{figure}
    \centering
    \includegraphics[width=8cm]{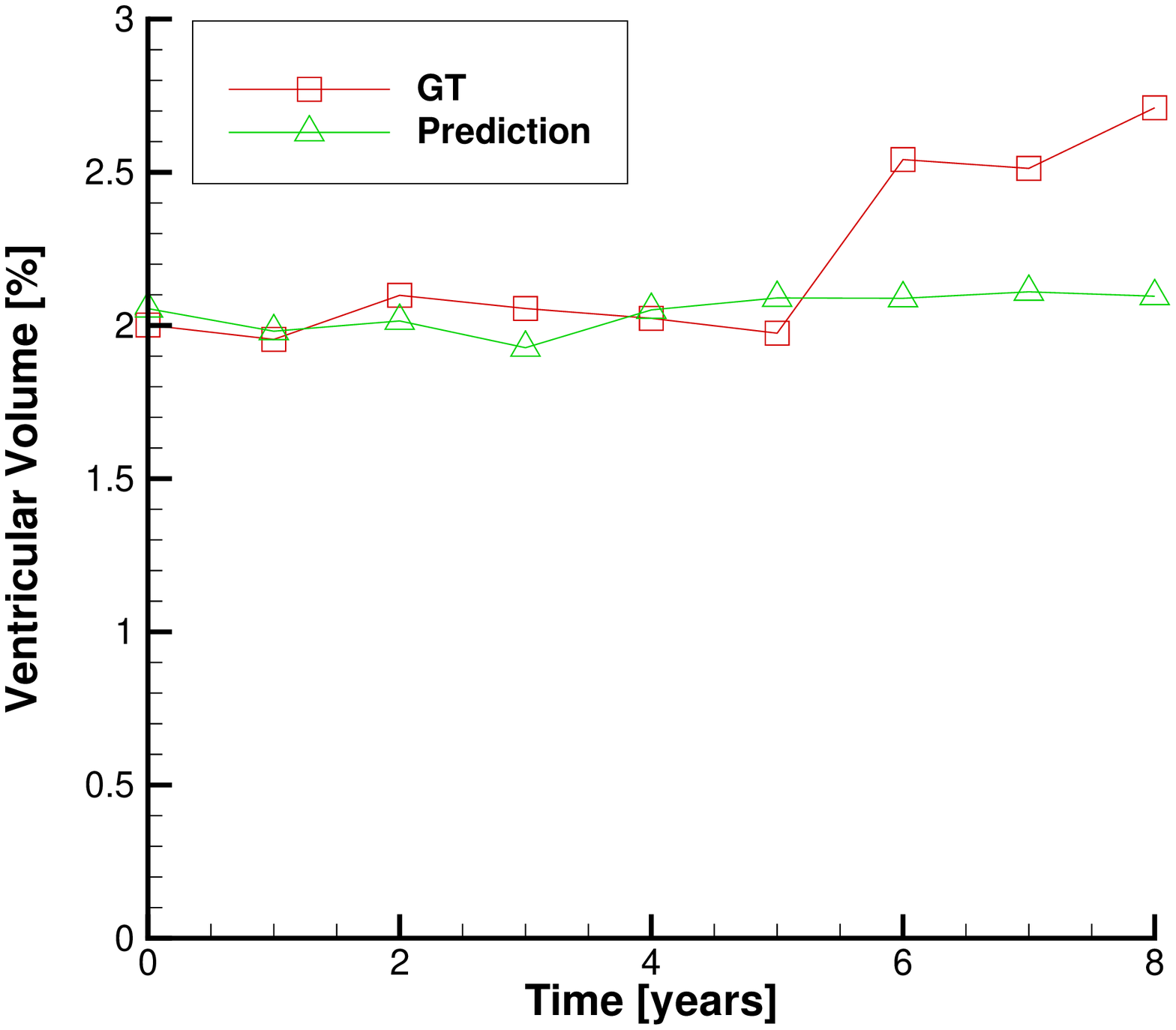}
    \caption{Ground truth and prediction results for ventricular volumes (case 2).}
    \label{fig:case2}
\end{figure}
\begin{figure}
    \centering
    \includegraphics[width=8cm]{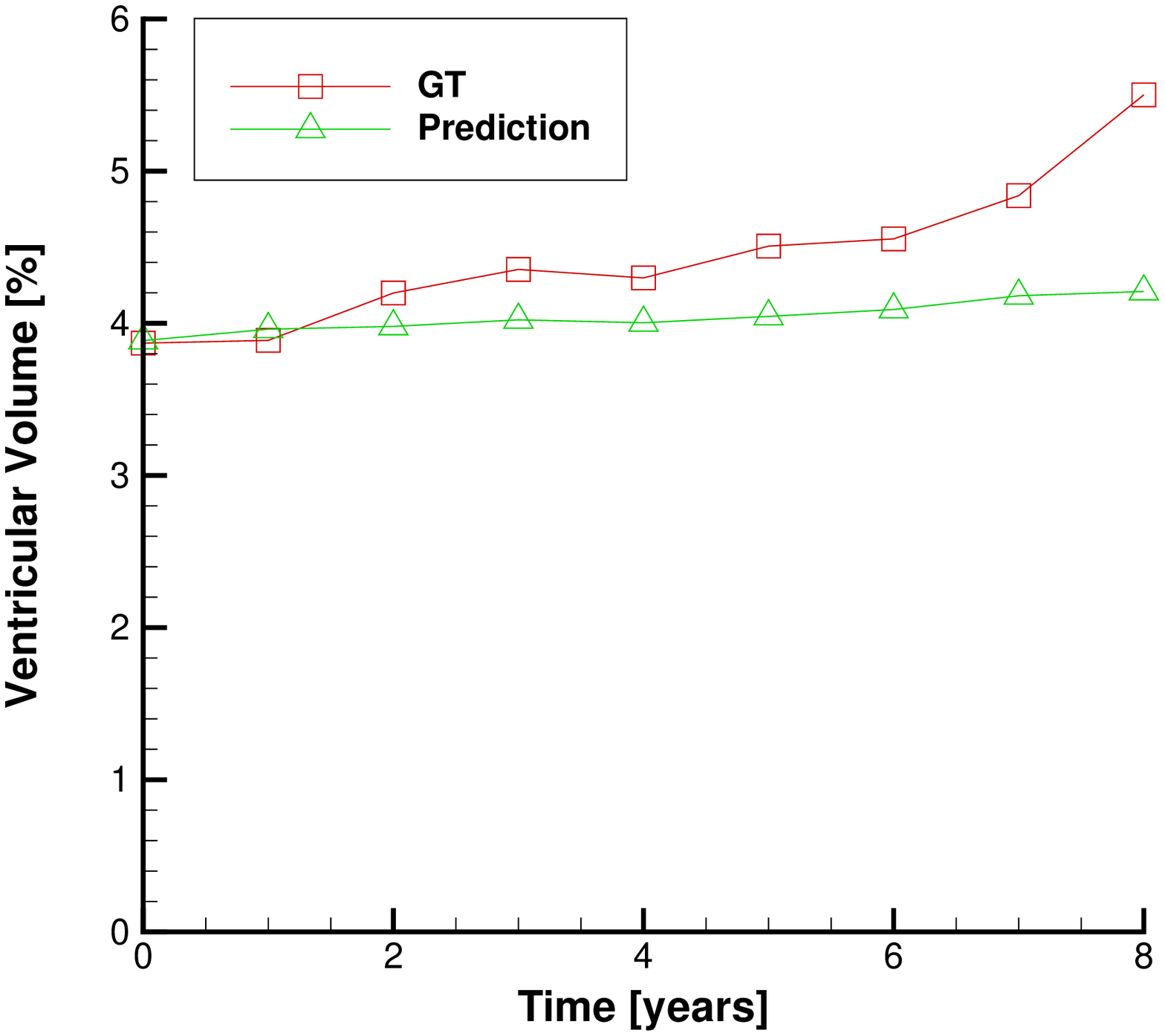}
    \caption{Ground truth and prediction results for ventricular volumes (case 3).}
    \label{fig:case3}
\end{figure}

Figures~\ref{fig:pred_case1}--\ref{fig:pred_case3} show the predicted aging brain over 8 years in three representative cross sections for three different subjects.
In these figures, ``G.T.'' means the ground truth images, ``Pred.'' means the predicted images obtained using our digital twin, and ``Diff.'' means differential images between the G.T. image of 0 years and the above Pred. image (the decrease in intensity is expressed in magenta and the increase is in blue).

Figures~\ref{fig:case1}--\ref{fig:case3} show the estimated ventricular volumes for the ground-truth and predicted images as a function of time.
We normalized these volumes with the total brain volume in the ground-truth image for each subject at 0 years.

\section{Discussion}
\subsection{Novelty and comparison}
We, for the first time, explicitly trained longitudinal conversion rules between volumetric medical images with a DGM.
Specifically, we used head CT images of healthy adults for the medical images to train our model and a full 3D flow-based DGM (3D-GLOW) for the model.

\begin{figure}
    \centering
    \includegraphics[width=12cm]{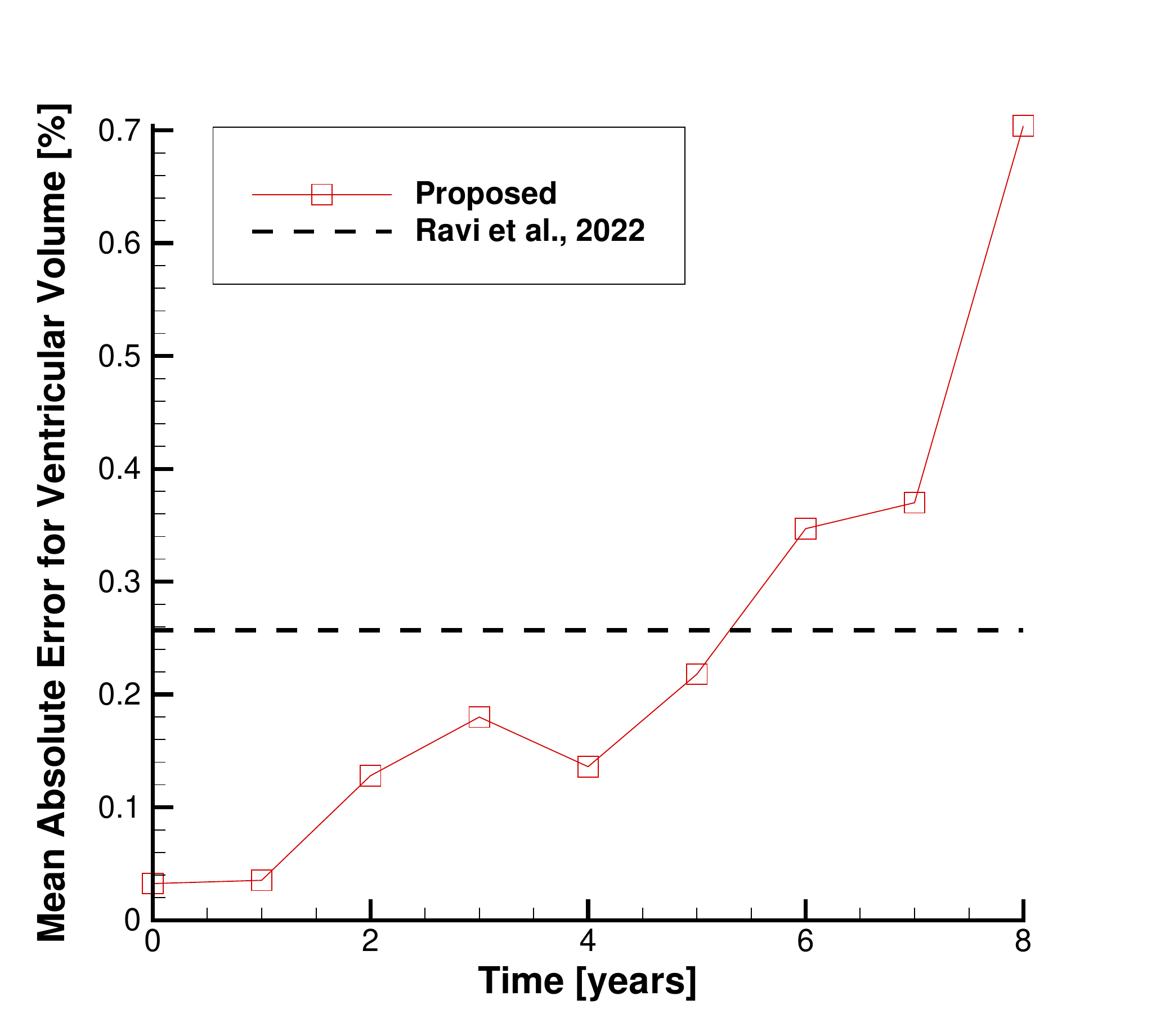}
    \caption{Comparison of the prediction accuracy in mean absolute error expressed in percentage of the total brain volume at 0 years between our proposed digital twin and the digital twin developed by Ravi et al. (2022).}
    \label{fig:averaged_mae_vs_time}
\end{figure}

Figure~\ref{fig:averaged_mae_vs_time} shows a comparison of the prediction accuracy in mean absolute error (averaged on three subjects) for ventricular volumes between our digital twin and the digital twin developed by Ravi et al. \cite{ravi2022degenerative}.
Ravi et al. did not refer to how many years they predicted aging of the brain to evaluate mean absolute error (MAE), but here, we explicitly show MAE for each year.
According to the graph, our digital twin has a prediction accuracy  equivalent or superior to that of the digital twin of Ravi et al. for ventricular volume if we assume the prediction term is within five years.  

\subsection{Modality}
In almost all of the previous studies including \cite{ravi2022degenerative}, MR images of the brain were used.
However, CT is less costly than MR imaging; hence, CT is often used in clinical practice.
Therefore, it is meaningful to predict brain atrophy quantitatively and objectively using CT images instead of MR images.
Our dataset comprises longitudinal 3D CT images (the acquisition interval is about one year).
Because CT images have a lower contrast--noise ratio (CNR) than MR images, our study is more challenging.
However, our results show at least equivalent prediction accuracy to previous studies, indicating the usefulness of our method.

\subsection{Clinical importance}
It is commonly known that (1) the hippocampus volume decreases in Alzheimer diseases, and (2) the frontotemporal lobe volume decreases in frontotemporal dementia.
However, we cannot evaluate the hippocampus, frontotemporal lobe, neocortex, and other regions with our digital twin.
This is because CT images have a lower CNR than MR images to extract the boundary between the gray matter and the white matter. 
We admit that our digital twin is inferior on this point.

Before our present study, there was no large-scale study in which aging-related atrophy of the whole brain was longitudinally examined using CT or MR images.
We, for the first time, examined errors in prediction accuracy for each year using a large-scale longitudinal 3D CT image dataset.
We plan to further evaluate prediction errors using MR images from the same subjects in our future study.

\subsection{Limitations}
From a one-point 3D head CT image, we successfully predicted the ventricular volume of an aged brain within six years, which reflects cerebral atrophy.
Still, after six years, we obtained a lower prediction accuracy.
The atrophy seemed gradual in the early stages of aging, but at a certain point of aging, it accelerated.
We found that our model cannot learn to specify a certain point, resulting in a lower prediction accuracy.
Furthermore, the present method can directly predict only a 3D head CT image after one year: we recursively obtained images after two years with this method.
With such sequential models, it may not be easy to model a certain point sufficiently.
We will address this difficulty in our future works.
For example, we will model 3D head CT images together with actual-age.
Alternatively, we will construct four-dimensional DGMs that can directly handle spatiotemporal volumetric images.

\section{Conclusion}
Before this study, no previous studies on aging prediction of an adult human body were carried out using DGMs with which longitudinal conversion rules of 3D medical images were explicitly trained.
We have successfully trained our digital twin for aging adult head CT images for the first time by adopting 3D-GLOW and longitudinally acquired 3D head CT images.
Our digital twin outperformed all previous digital twins in predicting ventricular volumes for relatively short terms, e.g., within five years, although we admit this is not a fair comparison.
To more fairly compare our results with those of previous studies, we plan to predict for longer years and quantify aging using MR images in our future study.

\section*{Acknowledgement}
This work was supported by JSPS KAKENHI Grant Number 21K18073.

 \bibliographystyle{elsarticle-num} 
 \bibliography{main}





\end{document}